
\input phyzzx
\nopagenumbers
\overfullrule=0pt
\baselineskip=18pt

\def\fun#1#2{\lower3.6pt\vbox{\baselineskip0pt\lineskip.9pt
  \ialign{$\mathsurround=0pt#1\hfil##\hfil$\crcr#2\crcr\sim\crcr}}}

\def\gap{\mathrel{\mathpalette\fun >}}
\def\order{{\cal O}}
\def\etal{{\it et al.}}

\def\Be{{$^7{\rm Be}(p,\gamma)^8{\rm B}$}}
\def\rar{\rightarrow}
\def\Hethree{{}^3{\rm He}({}^3{\rm He},2p){}^4{\rm He}}
\def\Hefour{{}^3{\rm He}(\alpha,\gamma){}^7{\rm Be}}
\def\Be{{}^7{\rm Be}(p,\gamma){}^8{\rm B}}
\def\Ni{{}^{14}{\rm N}(p,\gamma){}^{15}{\rm O}}


\Pubnum={IASSNS-HEP-93/22\cr IASSNS-AST 93/21}
\date={June 1993}
\titlepage
\title{Vacuum-Polarization Corrections to Solar-Fusion Rates}
\author{Marc Kamionkowski\foot{e-mail: kamion@guinness.ias.edu.}
and John N. Bahcall\foot{e-mail: bahcall@guinness.ias.edu}}
\address{School of Natural Sciences, Institute for Advanced
Study, Princeton, NJ 08540}
\abstract
The vacuum-polarization corrections to rates for nuclear-fusion
reactions in the $pp$ chain and in the CNO cycle are calculated.
For the reactions of particular importance to the solar-neutrino
problem, the $\Hethree$, $\Hefour$, $\Be$, and $\Ni$ reactions, we
find the magnitude of the effect to be less than 2\%.  The
effect of VP on all the other reaction rates is expected to be
of a similar order of magnitude.  We discuss how
these results affect the predicted fluxes of solar neutrinos.

\endpage
\baselineskip=18pt
\overfullrule=0pt
\pagenumber=2
\pagenumbers

\REF\parker{P. D. Parker and C. E. Rolfs, in {\it Solar Interior
and Atmosphere}, eds, A.~N.~Cox, W.~C.~Livingston, and
M.~S.~Matthews (University of Arizona Press, Tucson, 1992).}
\REF\bahcallbook{J. N. Bahcall, {\it Neutrino Astrophysics}
(Cambridge University Press, Cambridge, 1989).}
\REF\uehling{E. A. Uehling, {\sl Phys. Rev.} {\bf 48}, 55
(1935).}
\REF\gould{R. J. Gould, {\sl Astrophys. J.} {\bf 363}, 574 (1990).}
\REF\fowler{W. A. Fowler, {\sl Rev. Mod. Phys.} {\bf 56}, 149
(1984).}
\REF\krauss{A. Krauss \etal, {\sl Nucl. Phys.} {\bf A467}, 273
(1987).}
\REF\kavanagh{P. D. Parker and R. W. Kavanagh, {\sl Phys. Rev.}
{\bf 131}, 2578 (1963).}
\REF\kavanaghetal{R. W. Kavanagh \etal, {\sl Bull. Am. Phys.
Soc.} {\bf 14}, 1209 (1969).}
\REF\johnson{C. W. Johnson \etal, {\sl Astrophys. J.} {\bf 392},
320 (1992).}
\REF\schroder{U. Schr\"oder \etal, {\sl Nucl. Phys.} {\bf A467},
240 (1987).}
\REF\us{M. Kamionkowski and J. N. Bahcall, IASSNS-AST-93/90.  To
appear in {\sl Astrophys. J.} (1994).}

\FIG\figdeltavp{Plot of $\Delta(E)$ versus the center-of-mass
energy.  The solid line is for the $\Hethree$ reaction, the
short-dash line is for the $\Hefour$ reaction, the long-dash
line is for the $\Be$ reaction, and the dot-dash line is for the
$\Ni$ reaction.}


The cross sections for the nuclear reactions in the $pp$ chain
and in the CNO cycle are important for stellar-evolution
calculations of main-sequence stars [\parker]
and, in particular, for the calculated flux of solar neutrinos
[\bahcallbook].
The energies at which these cross sections are measured in the
laboratory are generally higher than the energies of interest in
the Sun.  The cross sections at solar energies are determined by
extrapolating the measurements to lower energies using the Gamow
penetration factor, which expresses the probability of quantum
tunneling through the Coulomb barrier.  In this paper, we
calculate the effect of vacuum-polarization (VP) corrections to the
electrostatic potential on the rates for nuclear reactions in the
$pp$ chain and in the CNO cycle.  We show that such corrections
lead to only small changes in the predicted flux of solar
neutrinos.

In addition to the Coulomb potential, $V_C=e^2/r$, there is an
additional contribution to the electrostatic potential, the
Uehling potential, $V_{VP}(r)$ [\uehling], which arises from
quantum corrections.  The complete electrostatic
potential for two nuclei with charges $Z_1$ and $Z_2$ separated
by a distance $r$ is
$$
V(r) = V_C + V_{VP}={Z_1 Z_2 e^2\over r}+ {Z_1 Z_2 e^2\over
r}\left({2\alpha I(r)\over 3\pi}\right),
\eqn\potential
$$
where $\alpha$ is the fine-structure constant, and
$$
I(r)=\int_1^\infty\, e^{-2m_e rx} \left(1+{1\over 2x^2}\right)
{(x^2-1)^{1/2} \over x^2}\,dx,
\eqn\Ieqn
$$
where $m_e$ is the electron mass.
The function $I(r)$ has the limiting forms
$$
I(r) = -\gamma - 5/6 - \ln(m_e r), \qquad {\rm for} \qquad m_e r\ll1,
\eqn\Ieqnone
$$
and
$$
I(r) = {3(2\pi)^{1/2} \over 4} {e^{-2 m_e r}\over (2 m_e r)^{3/2}}, \qquad
{\rm for} \qquad m_e r\gg1.
\eqn\Ieqntwo
$$
The function $I(r)$ has a logarithmic singularity for very small
radii, and exhibits an exponential falloff (arising from the
exchange of a virtual electron-positron pair) for $r\gap1/2m_e$.
The probability for tunneling through the electrostatic barrier
is affected by the presence of the Uehling
potential [\gould].

The energy dependence of a non-resonant fusion cross section is
ordinarily written (see, e.g., Refs.~[\bahcallbook] and [\fowler]),
$$
\sigma(E)\equiv {S(E) \over E} \Gamma^{(0)}(E),
\eqn\sigmaofE
$$
where $S(E)=S(0)+S'(0)E$ is a slowly varying function of $E$,
and $S(0)$ and $S'(0)$ are determined by fits to experimental
measurements of the cross section.  The quantity $\Gamma^{(0)}(E)$
is the usual Gamow penetration factor,
the probability of tunneling through the Coulomb barrier.
In practice, $S(E)$ is deduced by laboratory measurements of the
cross section at energies of order 100 keV to several MeV, and
the cross section is then extrapolated to energies, $\order(10\,
{\rm keV})$, typical of solar reactions, through Eq.~\sigmaofE.
In this paper, we use a WKB approximation similar to that used by
Gould [\gould] to estimate the vacuum-polarization
corrections to the Gamow penetration factor for any binary
reaction.

For a pure Coulomb potential, the
Gamow penetration factor for reacting nuclei with center-of-mass
energy $E$ is simply
$$
\eqalign{
\Gamma^{(0)}(E)&=\exp\left[ -{2\over\hbar}\int_0^b
\left[2\mu\left(V_C-E \right)\right]^{1/2} dr \right]\cr
&=\exp(-2 \pi \eta),\cr
}
\eqn\gamow
$$
where $b=Z_1 Z_2 e^2/E$ is the turning point radius, which is
determined by $V_C(b)=E$.
Here, $\eta=Z_1 Z_2 e^2/\hbar v$, where $v=(2E/\mu)^{1/2}$ is the relative
velocity of the incoming nuclei.  Also,
$\mu=M_p A_1 A_2/(A_1 + A_2)$ is the reduced mass of the
system, and $A_1$ and $A_2$ are the atomic mass numbers of the
reacting nuclei.

To include vacuum polarization in the cross section, we make the
substitution
$$
\Gamma^{(0)}(E) \rar \Gamma(E),
\eqn\substitution
$$
in Eq.~\sigmaofE, where $\Gamma(E)$ is the probability of
tunneling through the complete electrostatic potential.  Now,
$$
\Gamma(E)=\exp\left[
-{2\over\hbar}\int_0^{b_{VP}}\left[2\mu\left(V_C+V_{VP}-E
\right)\right]^{1/2} dr
\right]\equiv\Gamma^{(0)}(E)[1-\Delta(E)],
\eqn\barrier
$$
where the turning-point radius is now given by
$V_C(b_{VP})+V_{VP}(b_{VP})= E$, and can be written
$b_{VP}=b[1+2 \alpha I(b)/ 3\pi]$.  We expand the integral in
powers of the fine-structure constant, $\alpha$, and use the
result for $b_{VP}$ to find
$$
\Delta(E)= {4\alpha\eta \over 3\pi} \int_0^1 {I(bx)
\over \sqrt{x-x^2}} dx.
\eqn\deltavp
$$
The function $\Delta(E)$ is
plotted in Fig.~\figdeltavp\ for the $\Hethree$ (solid line),
$\Hefour$ (short-dash line), $\Be$ (long-dash line), and $\Ni$
(dot-dash line) reactions.
The function $\Delta(E)$
initially increases from its value at $E=0$ with increasing
energy, reaches a maximum at an energy $E_{max}$ and then
decreases roughly as $E^{-1/2}$ at
higher energies.  Although the lower limit of $\Delta(E)$ is
very conservatively given by 0 (at very large energies), the
lower limit of $\Delta(E)$ for energies at which measurements
are performed is actually not much smaller than $\Delta(E_{max})$.

The most probable energy of interaction for nuclei in the core of
the Sun is [\bahcallbook]
$$
E_\odot=1.22\,[Z_1^2 Z_2^2 (\mu/m_p) T_6^2]^{1/3}\, {\rm keV},
\eqn\solarenergy
$$
where $T_6$ is the temperature in units of $10^6$~K.  We take
$T_6=14$ (the temperature at which energy production is
maximized).
To self-consistently determine the effect of vacuum polarization
on reaction rates in the Sun, the VP correction must be included
in the analysis of the data from which the low-energy cross
sections are extrapolated.  This is done by fitting the data to
Eq.~\sigmaofE\ using $\Gamma$ instead of $\Gamma^{(0)}$.  The
cross section is then evaluated at energies typical of nuclear
reactions in the Sun with the $S(E)$ obtained and with
$\Gamma(E)$ rather than $\Gamma^{(0)}(E)$.  We then determine
the effect of VP by comparing this result for the cross section
with the standard result.

For the
$\Hethree$ reaction, we used data from Ref.~[\krauss] and found
that including VP self-consistently in the entire analysis
decreases the reaction rate by about 0.2\%.  For the $\Hefour$
reaction,
we used the Parker and Kavanagh data set [\kavanagh] and found that
VP decreases the reaction rate by 1.6\%.\foot{We are
grateful to P. Parker for providing the $\Hefour$ and $\Be$ data
sets.}  The decrease is
due to the fact that the data set is weighted at higher energies
where the VP correction is smaller than that at solar energies.
The magnitude of our result is also much smaller than 4.6\%,
which was obtained in Ref.~[\gould].  The discrepancy is due to
the fact that we included VP in fitting the
data, whereas Gould evaluated the correction to the cross
section at solar energies without correcting for the effect of
VP on the data.  In addition, our value of
$\Delta(E_\odot)=5.3\%$ differs from Gould's value (4.6\%)
because Gould evaluated the correction to the velocity factor in
the denominator of the WKB approximation to the wave function.
In the standard treatment, the Gamow penetration factor includes
only the leading exponential factor in the WKB approximation.
Therefore, for consistency with the standard experimental
procedure, we have not included the velocity
factor.  For the $\Be$ reaction, we used the data set of
Kavanagh \etal\ [\kavanaghetal], and
found that VP decreased the reaction rate by 0.1\%.  In order
to obtain this result, we used only data points at energies
below the resonance.  This is consistent with the treatment in
the most recent analysis of the $\Be$ $S(0)$ factor [\johnson].
The $\Ni$ reaction is the slowest process in the CNO cycle, so
the rate for production of neutrinos from ${}^{13}{\rm N}$ and
${}^{15}{\rm O}$ decays is controlled by the rate for this
reaction.  The best-estimate of the astrophysical $S$ factor for
this reaction comes from a data set made up of measurements
from 0.2 to 3.6 MeV [\schroder].  We used a simulated data set
made up of 18
uniformly spaced data points spread over this energy range and
found that VP decreased the solar reaction rate by 0.8\%.
For each of these four reactions, the effect of VP is much
smaller than the uncertainty in the measured value of $S(0)$.

Our data sets differ slightly from those used to obtain
the current best-estimates for the low-energy cross-section
factors.  Our results on the effect of VP would be
altered by negligible amounts if the exact data sets are used.

For the other reactions in the $pp$ chain and in the CNO cycle,
the effect of including VP in the entire analysis can
be crudely estimated by comparing the relative magnitudes of the
VP correction, $\Delta(E)$, at solar energies to that at
energies at which the measurements are performed.  Of course,
measurements are performed over some range of energies, and the
VP correction may change considerably over this range, but
conservative upper and lower bounds on the VP correction can be
provided.  We do so by comparing $\Delta(E)$
at solar energies with conservative estimates of the maximum and
minimum VP corrections to the data points.

In Table 1, we list the VP correction to the Gamow penetration
factor at solar energies, $\Delta(E_\odot)$, the most
probable energy of interaction, $E_\odot$, the maximum of
$\Delta(E_{max})=\Delta(E)$, and the energy, $E_{max}$ at which
this occurs.  In addition, we have listed the best-estimates,
$\delta_{VP}$, of the effect of VP on the rates for the
reactions that we have analyzed more carefully.  In all cases,
$\Delta(E_{max})$ is only slightly
larger than $\Delta(E_\odot)$, and $E_{max}>E_\odot$.
For each reaction, the effect of VP decreases the reaction rate
by no more than $\Delta(E_\odot)$ (which is always much less
than 6\%), and may increase the reaction rate by no more than
$\Delta(E_{max})-\Delta(E_\odot)$ (which is always less than 1\%).
The effect of inclusion of VP in the
entire analysis should always be much smaller than
$\Delta(E_\odot)$.  This is because $\Delta(E_{max})$
usually occurs near the lowest
energies at which measurements are performed, and
$\Delta(E)\sim E^{-1/2}$ remains non-negligible even at higher
energies (see Fig.~\figdeltavp).  This is illustrated by the
results for the reactions
that we have studied more carefully.  Therefore, VP should
generally have no more than an $\order(1\%)$ effect on
nuclear-reaction rates in the Sun.

The rate for the initial $p+p\rar {}^2{\rm H} +e^+ +\nu_e$
reaction is also affected by vacuum polarization
[\gould,\us].  The cross section is too small
for this reaction to be observed in the laboratory, so the rate
for the solar $pp$ reaction is calculated instead of
extrapolated from measurements at higher energies.  Therefore,
the effect of VP on the $pp$ rate is
determined in a different manner, and the result is that VP
decreases the rate by 0.6\% [\us].  We list
this best-estimate, as well $\Delta(E_\odot)$ and
$\Delta(E_{max})$, in the Table.

Our results imply that VP will have
little effect on the predicted flux of solar neutrinos.  The
flux of neutrinos from ${}^8{\rm B}$ decay depends most
sensitively on the nuclear-reaction rates.  The dependence of
the ${}^8{\rm B}$ neutrino flux may be written [\bahcallbook]
$$
\phi(^8{\rm B}) \propto S_{11}^{-2.6} S_{33}^{-0.40}
S_{34}^{0.81} S_{17}^{1.0},
\eqn\flux
$$
where $S_{11}$, $S_{33}$, $S_{34}$, and $S_{17}$ are the
low-energy cross-section factors for the $pp$, $\Hethree$,
$\Hefour$, and $\Be$ reactions, respectively.
Inserting our best-estimates for the effect of VP on the
low-energy cross-section factors, we find that inclusion of VP in the
analysis increases the predicted flux of ${}^8{\rm B}$ neutrinos
by only 0.2\%.

The dependence of the neutrino fluxes from the other reactions
in the $pp$ chain and CNO cycle on small changes in the
low-energy cross section factors are similarly determined
[\bahcallbook].  The flux of ${}^7{\rm Be}$ neutrinos will be
decreased by 0.7\% by VP corrections to the nuclear-reaction
rates; the neutrino fluxes from ${}^{13}{\rm N}$ and
${}^{15}{\rm O}$ decay will be increased by 0.8\% and 1.0\%,
respectively; and the flux of $pp$ neutrinos will be increased
by 0.02\%.  These result in an increase of 0.01 SNU in the
predicted event rate for the chlorine experiment and a decrease
of 0.11 SNU in the predicted event rate for the gallium
experiments.

In summary, we have evaluated the effect of vacuum
polarization on the determination of nuclear-fusion reaction
rates in the Sun.  We find that the magnitude of the VP
correction is never more than 2\% for the reactions that are
most important for solar-neutrino calculations.  The VP
correction has only a small effect on the calculated
solar-neutrino fluxes.

We acknowledge useful conversations with F.~Dyson and
R.~J.~Gould.  We are grateful to P. Parker for providing several
useful data sets.  We are also grateful to the referee for
pointing out an error in an earlier version of the manuscript.
MK was supported by
the Texas National Laboratory Research Commission, and by the
DOE through Grant No. DE-FG02-90ER40542.  JNB was supported by
the NSF through Grant No. PHY92-45317.

\refout
\figout
\vfil\eject

\vbox{\tabskip=0pt \offinterlineskip
\def\tablerule{\noalign{\hrule}}
\halign to450pt{\strut#& \vrule#\tabskip=1em plus2em&
  \hfil#&\vrule#&   \hfil#&\vrule#&   \hfil#&\vrule#&   \hfil#&\vrule#&
  \hfil#&\vrule#& \hfil#&\vrule#
  \tabskip=0pt\cr\tablerule
&&\multispan{11}\hfil VP Corrections to Nuclear-Reaction
Rates\hfil&\cr\tablerule
&&\omit\hidewidth Reaction\hidewidth&& \omit\hidewidth
$E_\odot$\hidewidth&& \omit\hidewidth $\Delta(E_\odot)$\hidewidth&&
\omit \hidewidth $E_{max}$\hidewidth&&
\omit\hidewidth $\Delta(E_{max})$\hidewidth&&
\omit\hidewidth $\delta_{VP}$\hidewidth & \cr\tablerule
&&\hfil $p+p \rar {}^2{\rm H}+e^+ +\nu_e$\hfil&&6&&1.4&&11&&1.5
&&-0.6&\cr\tablerule
&&\hfil $^3{\rm He}+^3{\rm
He}\rar\alpha+2p$\hfil&&20&&5.0&&46&&5.0 &&-0.2&\cr\tablerule
&&\hfil $^3{\rm He}+^4{\rm He}\rar^7{\rm
Be}+\gamma$\hfil&&21&&5.3&&46&&5.4 &&-1.6&\cr\tablerule
&&\hfil $^7{\rm Li}+p\rar2\alpha$\hfil&&14&&3.3&&31&&3.3 &&---&\cr\tablerule
&&\hfil $^7{\rm Be}+p\rar^8{\rm
B}+\gamma$\hfil&&17&&3.8&&46&&3.8 &&-0.1&\cr\tablerule
&&\hfil $^3{\rm He}+p\rar^4{\rm He}+e^+
+\nu_e$\hfil&&10&&2.5&&21&&2.5 &&---&\cr\tablerule
&&\hfil $^{12}{\rm C}(p,\gamma)^{13}{\rm
N}$\hfil&&23&&4.8&&66&&4.8 &&---&\cr\tablerule
&&\hfil $^{13}{\rm C}(p,\gamma)^{14}{\rm
N}$\hfil&&23&&4.8&&66&&4.9 &&---&\cr\tablerule
&&\hfil $^{14}{\rm N}(p,\gamma)^{15}{\rm
O}$\hfil&&25&&5.2&&76&&5.3 &&-0.8&\cr\tablerule
&&\hfil $^{15}{\rm N}(p,\gamma)^{16}{\rm
O}$\hfil&&25&&5.2&&76&&5.3 &&---&\cr\tablerule
&&\hfil $^{15}{\rm N}(p,\alpha)^{12}{\rm
C}$\hfil&&25&&5.2&&76&&5.3 &&---&\cr\tablerule
&&\hfil $^{16}{\rm O}(p,\gamma)^{17}{\rm
O}$\hfil&&28&&5.5&&91&&5.6 &&---&\cr\tablerule
}}
\item{\rm Table~1.}  Vacuum-polarization corrections to
nuclear-reaction rates.  $E_\odot$ and $E_{max}$ are in keV, and
$\Delta(E)$ and $\delta_{VP}$ are in \%.
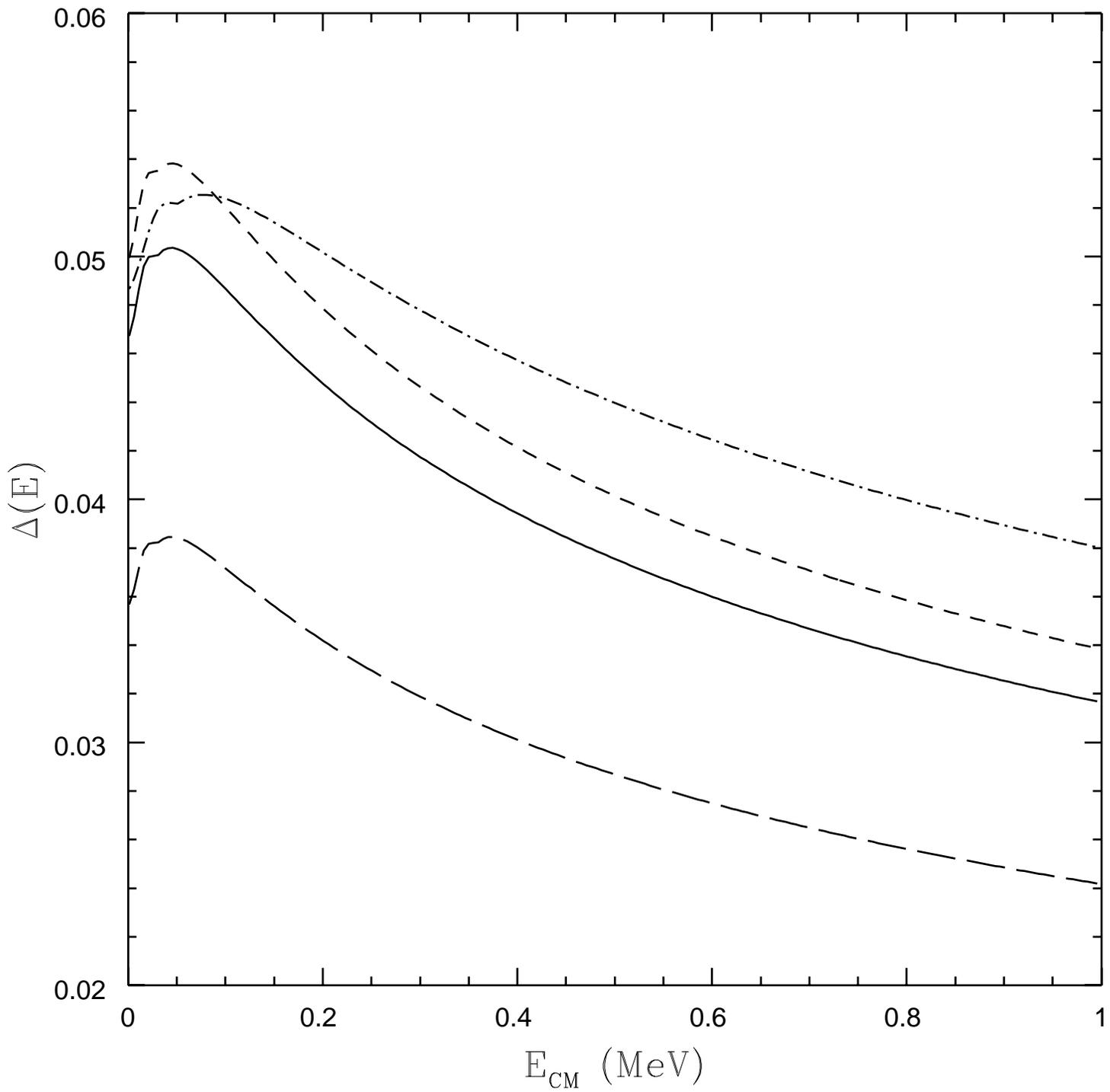
\end